\documentclass{aa}
\usepackage{graphicx}
\usepackage{natbib,multirow}
\bibpunct{(}{)}{;}{a}{}{,}
\begin{document}
   \title{Mid-Infrared observations of GRS\,1915+105 during plateau and 
flaring states}


   \author{Y. Fuchs
          \inst{1}
          \and
          I.F. Mirabel\inst{1,}\inst{2}
	  \and
	  A. Claret\inst{1}
          }

   \offprints{Y. Fuchs \\ \email{yfuchs@discovery.saclay.cea.fr}}
   \institute{Service d'Astrophysique, CEA/Saclay, Orme des Merisiers B\^at. 709, 91191 Gif-sur-Yvette, France
         \and
             Instituto de Astonom\'\i a y F\'\i sica del Espacio / CONICET, 
cc67, suc 28. 1428 Buenos Aires, Argentina
             }

   \date{Received 31 January 2003 / Accepted 21 March 2003}

   \abstract{ 

	We present mid-infrared (4--18\,$\mu$m) observations of 
	the microquasar GRS\,1915+105 obtained with ISOCAM, the
	camera on board the Infrared Space Observatory (ISO), in
	1996 April and 1997 October.
	The first observation probably
	occurred during a flaring event with oscillating synchrotron
	emission. The 1997 observation occurred a few days before a major
	relativistic ejection, during a plateau
	state of inverted-spectrum radio emission and hard
	quasi-stable X-ray emission. 
	The K-M giant donor star 
	in GRS\,1915+105 cannot account for the mid-IR emission and
	we discuss the possible additional components depending on
	two absorption laws. Thermal
	emission from dust 
	seems unlikely.
	The flat mid-IR spectrum obtained during the plateau state
	is likely to be synchrotron emission.  It would be
	the first evidence of the infrared extension of the radio
	synchrotron emission from the compact jets, although  
	optically thin free-free emission from an X-ray 
	driven-wind from the accretion disc cannot be
	excluded.

   }

   \maketitle
%

\section{Introduction}

	\object{GRS\,1915+105} is a transient hard X-ray source
	discovered in 1992 with Granat/WATCH  
	\citep{castro92,castro94}.  It is one of the brightest X-ray
	source in the Galaxy, highly variable in
	intensity and activity 
	\citep{harmon92,greiner96}. Intense flares are also
	observed in radio \citep{rodriguezmirabel93} and led to the
	discovery of superluminal ejections  with the VLA
	\citep{mirabelrodriguez94} i.e. structures moving
	with an apparent velocity higher than the light speed on the
	sky-plane. This first superluminal movement observed in our
	Galaxy, very similar to the ones observed in quasars like
	\object{3C\,279} \citep{cotton79}, confirmed the microquasar nature of
	GRS\,1915+105. The microquasar term refers to X-ray binaries
	showing radio (persistent or not) jets 
	imitating at much smaller scales the radio lobes of quasars
	and radio galaxies \citep{mirabel92,mirabelrodriguez99}.

	GRS\,1915+105 has a radio \citep{mirabelrodriguez93}
	and an IR \citep{mirabeletal93} counterpart, and is highly
	variable at all these wavelengths \citep{mirabeletal94}. 
	Because of the high interstellar extinction along the line of
	sight to GRS\,1915+105, the type of the companion star
	and thus the masses of the binary components were difficult
	to identify.
	According to \citet{greiner01} the donor star is of K-M\,III
	type corresponding to a late-type low mass
	($\sim$\,1--1.5\,M$_\odot$) giant star.  
	\citet{greiner01Nat} found the following binary
	parameters: orbital period \mbox{P$_\mathrm{orb}$ = 33.5 days,} mass
	function $f(M) = 9.5 \pm 3.0 $~M$_\odot$ which is the lower
	limit to the mass of the compact object, 
	using $M_\mathrm{d} = 1.2 \pm 0.2 $~M$_\odot$ for the 
	mass of the donor star
	and $\theta = 70^\circ \pm 2^\circ$ for the inclination
	angle implies $M_\mathrm{c} = 14 \pm 4 $~M$_\odot$ for the 
	mass of the compact object which is the most massive
	stellar black hole candidate.

\subsection{Multi-wavelength variability}

	GRS\,1915+105 is unique for its X-ray variability at all
	time scales from less than one second to several days
	\citep{greiner96}. RXTE observations of the source were
	classified in different classes (12 classes found by
	\citealt{belloni00} and a new class discovered by
	\citealt{kleinwolt02}) according to their flux and spectral
	properties and corresponding to transitions between 3
	basic states.
	Flaring events seen in radio, near-IR (K-band) and X-ray with
	various periodicity and intensity were classified in 3 classes
	by \citet{eikenberry00} as in the following.

	The first class is composed of the giant flares 
	that were observed in radio and X-ray
	\citep{mirabelrodriguez94,rodriguezmirabel99,fender99,dhawan00}
	and correspond to major ejection events followed as
	superluminal motions in radio images. These flares are
	characterized by very rapid rise time (less than 1 day) with
	an increase in the radio flux density by 1 or 2 orders of
	magnitude (reaching 600-1000\,mJy) and optically thin
	exponential decay lasting ten or so days \citep{foster96}.
	Continuous short period radio oscillations (between optically
	thin and thick states) can happen during the flux decrease
	\citep{fender99,fenderosc02}.

	These 20 to 40 minutes quasi-periodic oscillations form the
	second type of flares, first observed in radio with
	the Ryle Telescope at 15\,GHz \citep{pooley95,pooley96} and
	with the Very Large Array (VLA) at 3.6\,cm
	\citep{rodriguezmirabel97}. These oscillations are very common
	as shown by the Ryle Telescope monitoring of
	\citet{pooleyfender97} who associated them with soft X-ray
	dips on time-scale of $\sim$\,30 minutes. The latter were
	discovered by \citet{greiner96} with RXTE PCA observations and
	were interpreted by \citet{bellonimodel97,bellonidisc97} as
	the repeated disappearance and refilling of the inner
	accretion disc.
	Near-IR oscillations were also observed by
	\citet{fenderetal97} who suggested that they were the high
	frequency tail of radio synchrotron emission.  Simultaneous
	near-IR and X-ray observations by \citet{eikenberryinter98}
	revealed the correspondence between these X-ray
	dips/flares and near-IR (2.2\,$\mu$m K-band) flares consistent
	with synchrotron emission from ejected plasma bubbles or
	``baby jets'' undergoing adiabatic expansion losses.
	The synchrotron nature of these oscillations was confirmed by
	simultaneous near-IR/radio \citep{fenderpooley98} and
	near-IR/radio/X-ray \citep{mirabel98} observations of these
	$\sim$\,30~minutes dip/flare cycles, with wavelength-dependent
	time delays (radio-radio or radio-infrared).
	Recently \citet{fenderosc02} demonstrated that this time delay
	is variable between epochs, with a possible correlation with
	the oscillation amplitude, but they still cannot distinguish
	between models of discrete ejections (as the
	\citealt{vanderlaan66} model used by \citealt{mirabel98}) or
	shocks propagating along quasi-continuous flows
	\citep{blandford79,kaiser00}.


	The amplitudes of the quasi-periodic oscillations are about
	40--50~mJy in radio, or $\sim$\,70--80~mJy for the largest
	and $\sim$\,300 mJy in millimetre \citep{fenderpooley00}. In
	K-band the observed (not dereddened) amplitudes are
	$\sim$\,2--3~mJy \citep{fenderetal97,mirabel98,fenderpooley98} 
	$\sim$\,10~mJy \citep{eikenberryinter98,eikenberryspec98} 
	and $\sim$\,20~mJy \citep{fenderpooley00}.
	Finally, \citet{eikenberry00} presented a third type of flare:
	faint (submillijansky) IR (K-band) flares whose association
	with X-ray soft-dip/soft-flare behaviour is uncertain.\\


	Here we present the first mid-IR (4 to 18 $\mu$m) observations
	of GRS\,1915+105 thanks to the sensitivity of ISOCAM, the IR
	camera on board the Infrared Space Observatory (ISO).
	These observations were part of a general campaign studying
	the interaction of the high energy emission from X-ray
	binaries with the surrounding material. In particular, thermal
	emission from heated dust was expected as the presence of dust
	arround GRS\,1915+105 had been suggested by
	\citet{mirabel96} after observing a reddening at the time of a
	near-IR flare, and also by \citet{marti00} to explain the
	P~Cygni profile of a \ion{He}{i} line.
	In this paper we will first describe our ISOCAM observations
	and the difficulties in dereddening the data inherent to the
	mid-IR wavelength range. In section~\ref{obs96} and
	\ref{obs97} we show and interpret the results corresponding to
	two particular epochs in 1996 and 1997, when the source was in
	very different states: a possible flaring state in 1996
	and the {\it plateau} state preceeding a giant outburst with
	superluminal ejections in 1997 October. We also discuss the
	nature of this mid-IR emission. Section~\ref{concl} summarizes
	our conclusions.


        \begin{table*}[!htp]
        \caption{Log of the IR observations of GRS\,1915+105.
} 
	\label{tabobsGRS1915}
	\begin{minipage}{\hsize}
\begin{tabular}{l@{~}l@{~~}ll@{~}c@{~~}c@{~~}c}
\hline
\hline
date & MJD\footnote{Modified Julian Date = Julian Date $-$ 2400000.5} 
& timing & filter & wavelength & 
   Pixel Field & raster\\
 & & (UT) & & range & of View & size\\
\hline
28 April 1996 & 50201.45 & 10:55-11:03 & LW3 &  12--18\,$\mu$m & 
   $6''$$\times 6''$ & $ 2$$\times 2$\\
      	      &   & 11:03-11:10 & LW7 & 8.5--10.7\,$\mu$m &  
   $6''$$\times 6''$ &  $ 2$$\times 2$\\
      	      &   & 11:10-11:17 & LW1 &  4--5\,$\mu$m &  
   $6'' $$\times 6''$ & $ 2$$\times 2$\\
      	      &   & 11:17-11:25 & LW8 &  10.7--12\,$\mu$m &  
   $6'' $$\times 6''$ & $ 2$$\times 2$\\
28 April 1996 & 50201.48 & 11:26-11:35 & LW2 &  5--8.5\,$\mu$m &  
   $3'' $$\times 3''$ & $ 2$$\times 2$\\
 & & \multirow{3}{\jot}{\mbox{11:15-11:27}\,$\left\{ \begin{array}{l} \\ \\ \\ \end{array}\right.$} & J & 1.25\,$\pm$\,0.15\,$\mu$m &  
   0.125$''$$\times 0.125''$& 2$\times 2$\\
28 April 1996 & 50201.47 & &  H & 1.65\,$\pm$\,0.15\,$\mu$m &  
   0.125$''$$\times 0.125''$& 2$\times 2$\\
	& & &  K$_\mathrm{s}$ & 2.15\,$\pm$\,0.15\,$\mu$m &  
   0.125$''$$\times 0.125''$& 2$\times 2$\\
\hline
20 Oct 1997 & 50741.94 & 22:26-22:42 & LW10 &  8--15\,$\mu$m &  
   $1.5'' $$\times 1.5''$ & $ 2$$\times 2$\\
24 Oct 1997 & 50745.68 & 16:22-16:41 & LW3  & 12--18\,$\mu$m &  
   $3'' $$\times 3''$ & $ 3$$\times 3$\\
      	    &   & 16:41-17:00 & LW7 & 8.5--10.7\,$\mu$m &  
   $3'' $$\times 3''$ & $ 3$$\times 3$\\
      	    &   & 17:00-17:19 & LW9 & 14--16\,$\mu$m   
   & $3'' $$\times 3''$ & $ 3$$\times 3$\\
       	    &  & 17:19-17:38 & LW6 &  7--8.5\,$\mu$m   
   & $3'' $$\times 3''$ & $ 3$$\times 3$\\
       	    &  & 17:38-17:57 & LW8 & 10.7--12\,$\mu$m  &  
   $3'' $$\times 3''$ & $ 3$$\times 3$\\
       	    &  & 17:57-18:16 & LW4 &  5.5--6.5\,$\mu$m &  
   $3'' $$\times 3''$ & $ 3$$\times 3$\\
25 Oct 1997 & 50746.02 & 00:22-00:47 & LW2 &  5--8.5\,$\mu$m &  
   $1.5'' $$\times 1.5''$ & $ 3$$\times 3$\\
\hline
\end{tabular}
        \end{minipage}
        \end{table*}

        \begin{table*}[!htp]
        \caption{Infrared flux densities of GRS\,1915+105.
	$F_\mathrm{obs}$ is the observed flux directly measured on
	ISOCAM images, except for the J, H and K band fluxes which were
	reported from \citet{mahoney} and transformed into mJy units.
	$F_\mathrm{der}$ is the resulting flux after dereddening with two
	different laws: from \citet{lutz96} and from \citet{draine89}.}
	\label{tabfluxGRS1915}
\begin{tabular}{lc@{~}cc@{~}cc@{~}c}
\hline
\hline
Filter & \multicolumn{2}{c}{$F_\mathrm{obs}$ (mJy)} & \multicolumn{2}{c}{$F_\mathrm{der}$ Lutz (mJy)} & \multicolumn{2}{c}{$F_\mathrm{der}$ Draine (mJy)} \\
	& 1996 & 1997 & 1996 & 1997 & 1996 & 1997 \\
\hline
LW1 (4--5\,$\mu$m) &	11.2$\pm 3.4$ &      	      & 42.1$\pm 13$ &      	& 19.2$\pm 5.8$ &      \\
LW2 (5--8.5\,$\mu$m) &	5.39$\pm 1.6$ & 9.77$\pm 2.9$ & 17.4$\pm 5.2$ & 31.5$\pm 14$ & 7.72$\pm 2.3$ & 14.0$\pm 4.2$ \\
LW4 (5.5--6.5\,$\mu$m) &	     	      & 11.6$\pm 3.5$ & 	   & 37.1$\pm 11$ & 	     & 16.0$\pm 4.8$ \\
LW6 (7--8.5\,$\mu$m) &	     	      & 11.9$\pm 3.6$ & 	   & 38.4$\pm 12$ & 	     & 19.5$\pm 5.9$ \\
LW7 (8.5--10.7\,$\mu$m) &	1.53$\pm 0.5$ & 4.87$\pm 1.5$ & 7.37$\pm 2.2$ & 23.5$\pm 7.0$ & 7.26$\pm 2.2$ & 23.1$\pm 6.9$ \\
LW8 (10.7--12\,$\mu$m) &	3.54$\pm 1.1$ & 6.58$\pm 2.0$ & 10.3$\pm 3.1$ & 19.2$\pm 5.8$ & 10.8$\pm 3.2$ & 20.1$\pm 6.0$ \\
LW9 (14--16\,$\mu$m) &	     	      & 13.8$\pm 4.1$ & 	   & 23.9$\pm 7.2$ & 	     & 24.0$\pm 7.2$ \\
LW3 (12--18\,$\mu$m) &	5.34$\pm 1.6$ & 10.5$\pm 3.1$ & 9.25$\pm 2.8$ & 18.2$\pm 5.5$ & 9.29$\pm 2.8$ & 18.3$\pm 5.5$ \\
LW10 (8--15\,$\mu$m) &	    	      & 14.7$\pm 4.4$ &	  	   & 41.3$\pm 12$	&	     & 36.0$\pm 11$ \\
J (1.25\,$\pm$\,0.15\,$\mu$m) &	0.688$\pm 0.04$ &  & 121.0$\pm 7.8$ & & 135.2$\pm 8.7$ & \\
H (1.65\,$\pm$\,0.15\,$\mu$m) &	3.14$\pm 0.14$ &   & 78.3$\pm 3.6$ & & 77.1$\pm 3.6$ & \\
K$_\mathrm{s}$ (2.15\,$\pm$\,0.15\,$\mu$m) & 6.98$\pm 0.26$ &   & 52.7$\pm 1.9$ & & 51.8$\pm 1.9$ & \\
\hline
\end{tabular}
        \end{table*}

\section{Observations}

   \subsection{ISOCAM}
	Mid-IR observations of GRS\,1915+105 were carried out with the
	ISOCAM camera \citep{cesarsky96} on board the Infrared
	Space Observatory (ISO) satellite mission as part of the
	Guaranteed and Open Time programmes.
	Table~\ref{tabobsGRS1915} shows the log of the
	observations for the two epochs in 1996 April and 
	1997 October.

   \subsubsection{Data reduction}

	The ISOCAM data were reduced with the Cam Interactive Analysis
	software (CIA, \citealt{Delanay00}) version 4.0, following the
	standard processing outlined in \citet{starck99}.  First a
	dark correction was applied, then a de-glitching to remove
	cosmic ray hits, followed by a transient correction
	to take into account memory effects. Pixels showing remnants of
	these effects were masked as well as side pixels insufficiently
	lit. The jitter correction was applied for
	$1.5''\times 1.5''$ resolution images. The flat-field
	correction used the automatic evaluation except for the
	LW1 filter where the calibration flat field was used.  Then
	individual images were combined into the final raster map
	and finally the pixel values were converted into milli-Jansky
	flux densities. No colour correction was applied.

   \subsubsection{Images and photometry} 
	
	GRS\,1915+105 appears on the ISOCAM images as a very faint
	point source, at the limit of the ISOCAM detection for a few
	filters in 1996.  No extended emission was found, but the
	0.3--0.6$''$ long extensions seen by \citet{samseckart96} at
	2.2\,$\mu$m in 1995 July are far too small for ISOCAM spatial
	resolution. Note that these near-IR ``jets'' were suggested to
	be of synchrotron nature and variable, but their presence was
	never confirmed. \citet{samseckartiauc96} noted the
	disappearence of the jet since their previous observation and
	\citet{eikenberryfazio97} found no evidence of near-IR
	extended emission.
 
	Achieving the photometry of GRS\,1915+105 was a delicate task
	as it is located close to a very bright mid-IR source (J2000
	coordinates $\alpha = 19^h 15^m 11.27^s$ and $\delta =
	10^\circ 56' 27''$). This source is the star visible in
	Figure~2 of \citet{mirabeletal94} at the South-West of the
	ROSAT error circle on this R-band image, and in
	\citet{fenderetal97} Figure~1 (next to star ``B'') where it
	appears as already quite bright in the K-band. The $\sim 18''$
	distance (center to center) between this star and
	GRS\,1915+105 corresponds to only 12, 6 and 3 pixels in our
	ISOCAM images depending on the spatial resolution.
	This star is very bright in all the ISOCAM images and its point
	spread function certainly overlap the one of
	GRS\,1915+105. Thus we measured the integrated
	flux inside a tight circle around the source peak and we used
	an aperture correction to obtain the observed flux $F_\mathrm{obs}$.
	The photometry accuracy was estimated to $\sim$\,30\% mainly because
	of measurement errors due to the presence of this bright star.

	As shown in Table~\ref{tabobsGRS1915}, GRS\,1915+105 was
	observed at different epochs with different filters and
	different spatial resolutions which raises doubt about whether the
	flux variability is real or not.  Comparison with the bright
	source shows that its flux is stable on all our observations
	(within the error bars) and so the variability of the
	flux densities measured for GRS\,1915+105 between 1996
	and 1997 shown in Table~\ref{tabfluxGRS1915} is real.

   \subsection{Near-IR values}

	We have looked for simultaneous observations of GRS\,1915+105
	with our ISOCAM ones and we have found that \citet{mahoney}
	observed the source in the J, H and K$_\mathrm{s}$
	(2.15\,$\mu$m) bands on April 28 1996 at the same time
	as ISOCAM LW1, LW8 and LW2 filters as shown in
	Table~\ref{tabobsGRS1915}.
	In this table the J, H and K$_\mathrm{s}$ band fluxes were
	reported from \citet{mahoney} after conversion from
	magnitude into milliJansky units according to \citet{zombeck}
	reference fluxes. They were dereddened using the \citet{lutz96}
	and \citet{draine89} laws as described below in
	section~\ref{secder}.  These laws give at these 3 near-IR
	bands very similar results to the more commonly used
	\citet{cardelli89} law, so as they were used in the mid-IR
	range we kept them in the near-IR for coherence.
	Error bars on these bands were calculated by adding
	observational errors given by \citet{mahoney} and errors
	caused by dereddening, about 10\% for J and 2\% for H and
	K$_\mathrm{s}$, so they are respectively 12\%, 5\% and 4\%.

	We have also looked for the maximum and minimum published values in 
	the J, H and K bands. The minima come from \citet{chaty96},
	the maxima come from \citet{mahoney}, \citet{mirabel96} and
	\citet{chaty96} for the J, H and K-bands. We
	transformed these fluxes into milliJansky units and dereddened
	them.

\section{The dereddening problem}

\subsection{Distance and visible absorption}

	The distance of GRS\,1915+105 is not known
	precisely. Observation of the superluminal ejecta enables us to
	constrain the maximum distance. \citet{fender99} calculated
	their relativistic proper motion and found \mbox{$11.2 \pm 0.8$\,kpc}
	as an upper limit. They also discussed the minimum distance,
	which is $> 6.6$\,kpc.
We adopt the value of 11.2\,kpc.

	At such a distance and because GRS\,1915+105 is located in the
	galactic plane with $\alpha$ = 19$^{h}$ 15$^{m}$ 11.5494$^{s}$
	and $\delta$ = +10$^{\circ}$ 56$'$ 44.758$''$ in J2000
	coordinates \citep{dhawan00} corresponding to $l=
	45.3657^{\circ}$ and $b= -0.219^{\circ}$, the visible
	radiation from this source is heavily absorbed by the material
	along the line of sight. The optical extinction $A_V$ was roughly
	evaluated to be $A_V \sim 30$ magnitudes by \citet{mirabeletal94} and
	$A_V \simeq 28$\,mag by \citet{boer96}. \citet{chaty96}
	derived from millimeter $^{12}$CO\,(J=$1-0$) observations a total
	hydrogen column density $N_\mathrm{H} = 4.7 \pm 0.2 \times 10^{22}$
	cm$^{-2}$ along the line of sight corresponding to $A_V = 26.5
	\pm 1$ mag. This value was recently corrected by
	\citet[ in\ prep.]{chapuis03} who find $N_\mathrm{H} = 3.5 \pm 0.3
	\times 10^{22}$ cm$^{-2}$ and thus $A_V = 19.6 \pm 1.7$
	mag. In this article we adopt \mbox{$A_V = 20$ mag.}
 
	Note that this absorption coefficient is the main parameter
	used to deredden optical and infrared fluxes. With $A_V=20$
	and using either the \citet{riekelebo85} or the \citet{cardelli89}  
	law one finds $A_K=2.2$ (see also \citealt[ in\ prep.]{chapuis03}) and
	not the 3.3 value, considered as a possible significant overestimate by \citet{fenderpooley98} but widely used in previous articles with
	near-IR observations. Then for example the observed
	$\sim$\,3\,mJy K-band oscillations amplitude leads when dereddened
	not to $\sim$\,60 mJy but to $\sim$\,20 mJy which is quite
	different from the radio oscillation amplitude.  
 	As a consequence the K-band flux in general is not as high as
 	previously cited and thus the radio to IR spectrum is not as
 	flat \citep{fenderetal97,fenderpooley00}, although the simultaneous radio and millimetre observations \citep{fenderpooley00} in which there are no dereddening uncertainties show a quite flat spectrum.
However the
 	flat spectrum argument may not be used  against
 	the \citet{vanderlaan66} model of expanding clouds of plasma.

   \begin{figure}[!tp]
	\includegraphics[width=9.0cm]{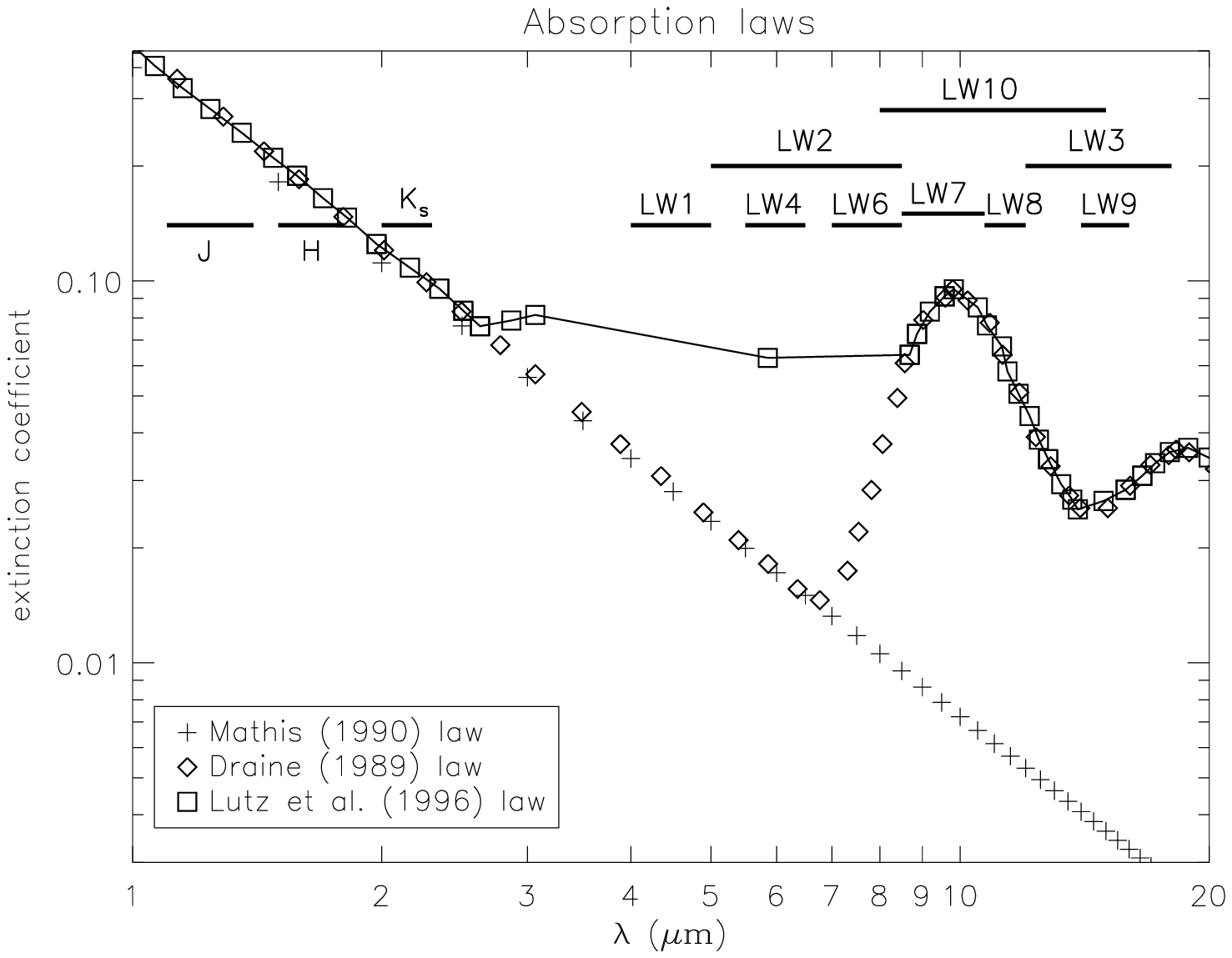}
	\caption{Comparison of the near-IR absorption law by
	\citet{mathis90} where $A_\lambda \propto \lambda^{-1.7}$
	(plus sign) with \cite{draine89} and \citet{lutz96} laws
	(respectively diamonds and squares) in the mid-IR. The
	extinction coefficient is defined by $A_\lambda / A_V$ where
	$A_V$ is the exctinction in the optical V (0.55\,$\mu$m) band
	and $A_\lambda$ is the exctinction at wavelength
	$\lambda$. The near-IR and ISOCAM filter wavelength ranges used in
	this article are overplotted.}
	\label{figabs}
	\end{figure}

   \subsection{Dereddening}
		\label{secder}

	Mid-IR data need specific dereddening as the absorption law is
	highly wavelength dependent and irregular in this
	wavelength range compared to the near-IR range. 
	The \citet{cardelli89} or \citet{mathis90} laws are no
	longer valid for $\lambda > 2.5\,\mu$m (so in the
	mid-IR) as they totally ignore typical features such as the ones
	due to silicates.  The Galactic interstellar medium was
	studied separately by \citet{draine89} and \citet{lutz96} who
	found two different absorption laws, particularly in the
	2.5--8.5~$\mu$m ($3.5-12 \times 10^4$ GHz) range. 
	Fig.~\ref{figabs} shows the comparison between the \citet{mathis90},
	\citet{draine89} and \citet{lutz96} laws.
	According to Anthony Jones \citetext{priv.\ comm.} the
	\citet{lutz96} law applies for the galactic center, for
	compact \ion{H}{ii} regions and for particular sources
	surrounded by high carbon density regions. The
	\citet{draine89} law is supposed to apply in more diffuse
	regions so for the majority of Galactic sources. GRS\,1915+105
	is not in the Galactic center region but it is in the dense
	Sagittarius arm, and we do not know its chemical
	environment. Thus, we could not choose
	between one or other law, so we have dereddened the flux
	densities with both laws and have studied each case in
	the following sections.

	Note that what we call dereddened flux $F_\mathrm{der}$ is linked to
	the observed flux $F_\mathrm{obs}$ by the formula:
	\mbox{$F_\mathrm{der}\,=\,F_\mathrm{obs}\,\times\,10^{+0.4A_{\lambda}}$}
	where $A_{\lambda}$ is the extinction at wavelength $\lambda$,
	calculated as $A_V \times$ the extinction
	coefficient at $\lambda$ given by either \citet{draine89} or
	\citet{lutz96} law.

	Dereddening introduces an error of about 5\% into the flux
	density so it is negligible compared to the 30\% 
	error due to the photometry. Thus the total error on ISOCAM
	dereddened flux densities is $\sim$\,30\% of the flux
	densities (see Table \ref{tabfluxGRS1915}).


\section{A high/soft state in 1996 ?}
\label{obs96}

	\subsection{State of the source}
	We observed GRS\,1915+105 with ISOCAM for the first time on 
	April 28 1996 (MJD~50201). Observed and dereddened fluxes are given in
	Table~\ref{tabfluxGRS1915}.

   \begin{figure}[!tp]
	\hspace*{-0.4cm} \includegraphics[width=9.0cm]{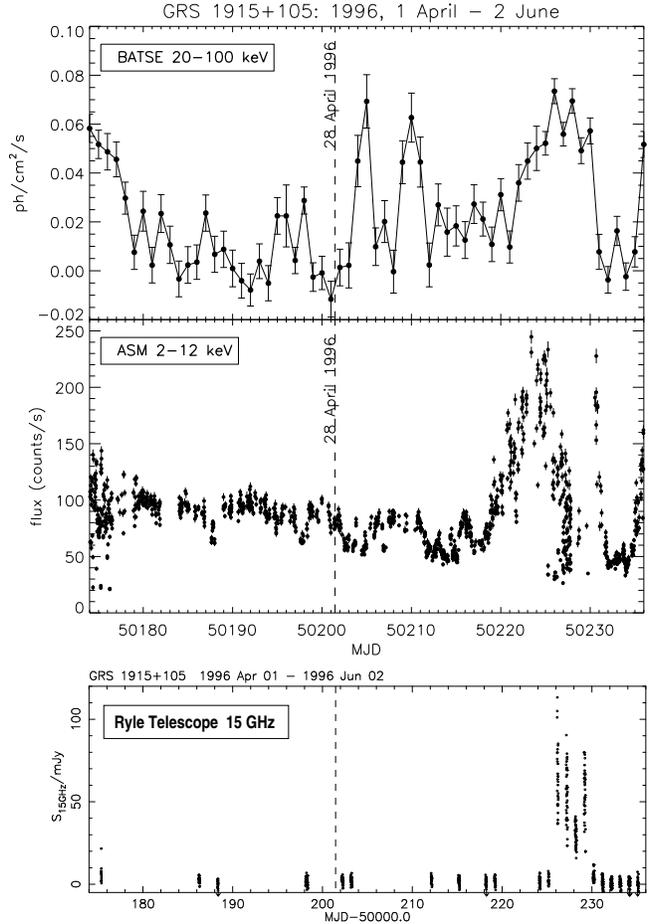}
	\caption{X-ray and radio flux monitoring of GRS\,1915+105
	around April 28 1996 (MJD 50201). The dotted line marks the
	begining of our ISOCAM observation on that day. The upper
	panels show the photon flux measured by BATSE in the
	$20-100$~kev band and the integrated count rate in the
	$2-12$~keV band by the Rossi X-ray Timing Explorer All-Sky
	Monitor (RXTE ASM). The lower panel shows the corresponding
	flux density measured at 15\,GHz with the Ryle Telescope 
	(plot kindly provided by Guy Pooley).}  \label{figetat96}
	\end{figure}

   \begin{figure*}[!htp]
	\includegraphics[width=8.8cm]{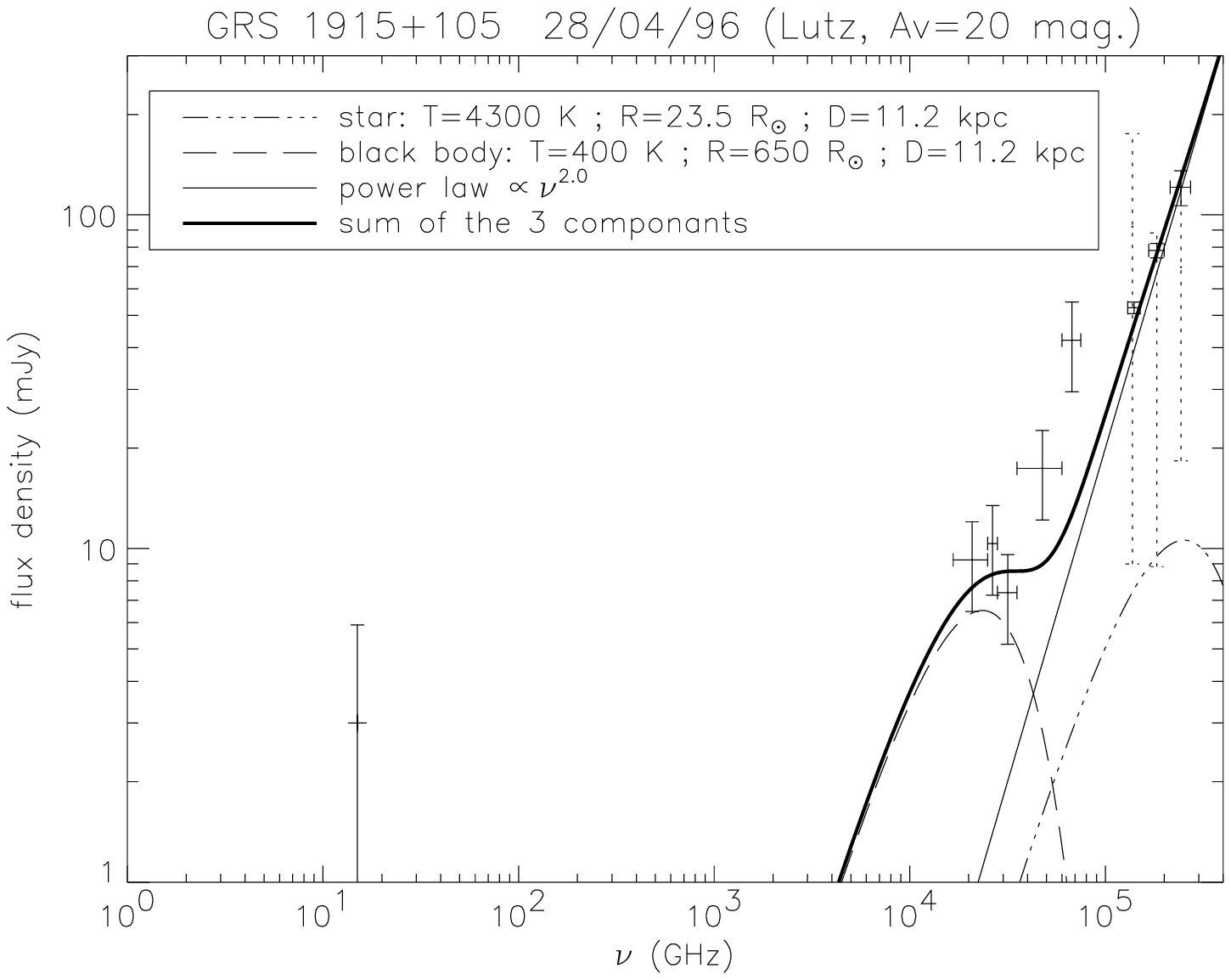}
	\includegraphics[width=8.8cm]{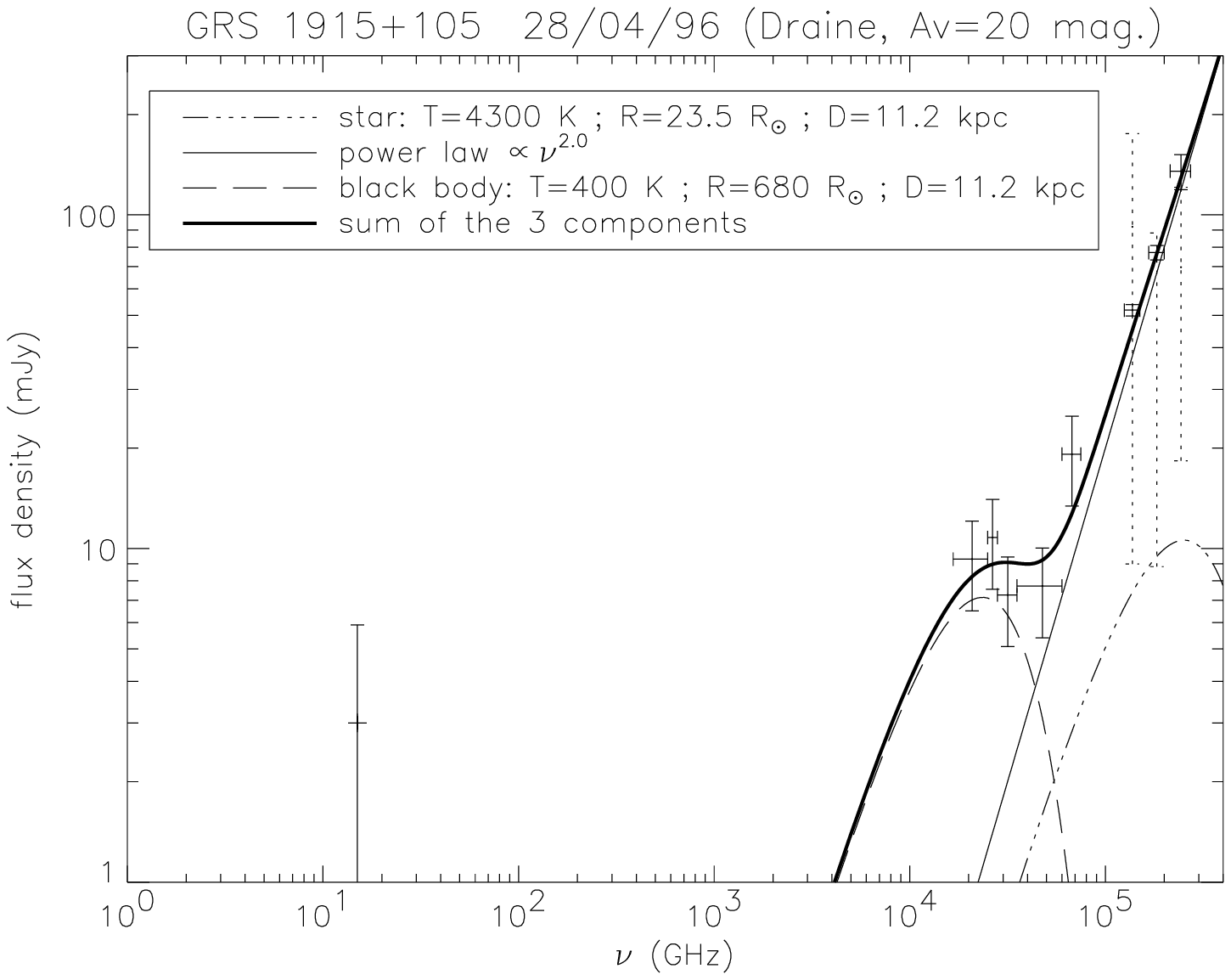}
   	\caption{IR spectrum of GRS\,1915+105 in 1996 April 28,
   	dereddened with the (left) \citet{lutz96} and (right)
   	\citet{draine89} laws. The dotted lines correspond to the minima
   	and maxima flux densities published in the 
	J (1.25\,$\mu$m $= 2.4 \times 10^5$ GHz), 
	H (1.65\,$\mu$m $= 1.81 \times 10^5$ GHz) and K
   	(2.2\,$\mu$m $= 1.36 \times 10^5$ GHz) bands. ISOCAM filters
   	are, from left to right: LW3 ($1.67-2.5 \times 10^4$ GHz), 
	LW8 ($2.5-2.8 \times 10^4$ GHz), LW7 ($2.8-3.53 \times 10^4$ GHz), 
	LW2 ($3.53-6.0 \times 10^4$ GHz), LW1 ($6.0-7.5 \times 10^4$ GHz) 
   	(see Table~\ref{tabobsGRS1915}). The low radio level observed with
   	the Ryle Telescope at 15\,GHz 2 days before and one day after
   	the ISOCAM observations was added. Tentative fits of the IR
   	flux densities are plotted as indicated in the legend (see details in
   	the text).}
	\label{figISO96}
    \end{figure*}

	Because of the high variability of GRS\,1915+105, it is
	important to know the state of this source for data
	interpretation. The radio and X-ray monitorings of
	Fig.~\ref{figetat96} illustrate the changes in the flux
	densities.  The quick-look results provided by the ASM/RXTE
	team shows the integrated count rate of GRS\,1915+105 over the
	2--12~keV band. The source behaviour was quite variable
	during the month preceeding and following our IR observations
	(see also the light curve in Fig.~4 of
	\citealt{pooleyfender97}). On MJD~50201, GRS\,1915+105 showed a
	relatively high soft X-ray flux around 80\,counts/s.  The
	corresponding BATSE occultation data showed no detectable
	hard X-ray flux in the 20--100~keV range on the same day. Thus,
	the state of GRS\,1915+105 during our IR observations is
	moderately high and moderately soft.

	There was no simultaneous radio observation of this period as
	the Green Bank Interferometer (GBI) was not operating. The
	Ryle Telescope (RT) was observing GRS1915+105 at 15 GHz but
	not continuously as shown in Fig.~\ref{figetat96} (RT
	plot kindly provided by Guy Pooley). There were
	observations taken on MJD 50199 and MJD 50202 i.e. about 2 days
	before and one day after our ISOCAM observation (see also
	\citealt{pooleyfender97}). They show a very low flux level of
	about $3\pm 3$~mJy classified as a detection \citetext{Pooley\
	priv.\ comm.}  so the source was in a low radio state through
	the period of our observation.

	\subsection{Results}

	In order to understand the nature of the IR emission from
	GRS\,1915+105, we have tried to fit it using simple
	models. X-ray binaries gather many sources of IR radiation:
	the donor star, the external edge of the accretion disc,
	thermal reprocessing of the X-ray disc flux, the basis of
	possible compact jets, heated dust around the system...

	The most obvious IR source is the donor star. It was
	identified by \citet{greiner01} as a K-M giant star with a
	temperature estimate of $\sim 4800^{+200}_{-500}$~K and a
	magnitude of K=14.5--15.0 uncorrected for extinction.
	\citet{greiner01Nat} estimated the Roche lobe size to be $21 \pm
	4$~R$_\odot$. We have fitted the star contribution with
	a black body at a distance of 11.2\,kpc, with a temperature T
	and radius R. We have chosen the lowest temperature
	$\mathrm{T}=4300$\,K since it corresponds to the most
	displaced spectral curve toward the mid-IR range. We have 
	found the black body radius R=19--23.5~R$_\odot$ by
	comparing the resulting black body flux in the
	K-band with the dereddened K magnitude.  Finally we have
	chosen $\mathrm{R}=23.5$~R$_\odot$ in order to model the 
	star contribution with the highest flux density. 
	This black body is plotted with the dashed-dotted line in
	Fig.~\ref{figISO96} where it is clear that this contribution
	is faint in both near-IR and mid-IR range.

	Thus other contributions are needed to explain the IR
	emission. 
	Near-IR fluxes can be
	approximatively fitted by a power law ($F_\nu \propto
	\nu^\alpha$) with a spectral index $\alpha = 2$ (thin solid 
	line in Fig.~\ref{figISO96}) but its interpretation is
	questionable. It may correspond to optically thick free-free
	emission but it would be observed at unusually high
	frequencies. This power law may also be due to the external
	edge of the accretion disc at a temperature of $\sim 8000$~K
	(and radius a $\sim 35$~R$_\odot$), it would be the low energy
	tail of the multi-black body model commonly used to account for the
	majority of the soft X-ray emission in the high/soft state.
		However, as shown in Fig.~\ref{figISO96} this power law
	emission is not sufficient to explain the mid-IR emission.
	Thus an additional component is needed
	in the mid-IR range. This component may be thermal emission
	from dust as plotted with the long-dashed line in
	Fig.~\ref{figISO96} using a simple black body model with
	\mbox{$\mathrm{T}=400$~K} and $\mathrm{R}=650$~R$_\odot$ (with
	\citealt{lutz96}) or $\mathrm{R}=680$~R$_\odot$ (with
	\citealt{draine89}).

	The sum of the 3 components is plotted with a thick solid line
	in Fig.~\ref{figISO96}, but the result is moderately
	satisfactory for flux densities dereddened with the
	\citet{draine89} law and not satisfactory for flux 
	densities dereddened with the \citet{lutz96} law.

	\subsection{Interpretation: a possible flaring event ?}

	This non-satisfactory fit of the April 28 1996 (MJD 50201)
	observations may be due to a flare occuring at the same moment
	in GRS\,1915+105.

	\citet{pooleyfender97} described MJD 50135--50220 as a period
	of varying degree of activity but with no significant radio
	emission apart from one small event. Their figure~4 shows
	indeed a quiet radio period but with a rather sparse coverage
	with the Ryle Telescope at 15 GHz.  Nevertheless, it is
	possible that a brief flare or a brief and isolated flaring
	period had occurred just during our ISOCAM observation since
	the J-band flux measured by \citet{mahoney} is the highest
	ever reported for this wavelength filter and since very short flares
	can occur as the one seen just before MJD 50440 in Fig.~4 of
	\citet{pooleyfender97}. This possible brief IR flare 
	could have occurred simultaneously in the radio wavelengths but as it
	lasted less than one day, no trace is left on the MJD 50202
	Ryle Telescope observation.

	This case assumes that the near-IR observations of
	\citet{mahoney} must no longer be considered as simultaneous. 
	Indeed it seems unlikely when one compares
	the near-IR observations which lasted 12 minutes with 4
	minutes for each band \citep{mahoney} with the typical 30
	minute duration of the quasi-periodic oscillations of
	GRS\,1915+105. For example, \citet{fenderetal97} show
	oscillations with a flare duration of $\sim 15$ minutes,
	so 3 observations of 4 minutes each taken during this kind of
	oscillation can lead to very different flux levels and
	consequently to a big slope, as the one between the J, H
	and K$_\mathrm{s}$ bands in Fig.~\ref{figISO96}. The
	extrapolation of this slope is not consistent with a 
	detectable radio signal and this would be completely
	inconsistent with all the past recorded radio flares
	corresponding to high mean flux densities of the order
	of 25 to 50 mJy.

	This possible flaring event is not identified as
	a peculiar variability of the flux evolution in each
	successive image taken with ISOCAM, perhaps because of the slow
	response of the mid-IR detectors emphasized by the weakness of
	the source. Thus flux densities measured in the different ISOCAM filters during this flare have
	to be considered as time averages (about 7 minutes per
	filter). As a consequence, the resulting spectral energy
	distribution does not represent a unique spectrum of
	GRS\,1915+105 and one should not rely too heavily on its
	modelisation. Near-IR oscillations observed as $\sim$\,30
	minutes dip/flare cycles in the past were considered as
	synchrotron emission (see section~1.1.).

	Another case to take into account is if no flare occurred
	during our ISOCAM observation on MJD 50201. Then one has to
	explain why the J-band flux observed by \citet{mahoney} is the
	highest ever recorded. A possible explanation would be that
	the accretion disc had been particularly bright at that time,
	leading to high near-IR flux densities coming from the
	outer edge of the accretion disc, but it is unlikely as the
	RXTE/ASM flux was not exceptionally high.

	Thus, the only reliable argument is that the donor star
	cannot account for most of the near and mid-IR emission of the
	source. One or several additional components are needed to
	explain this 1996 observation, such as synchrotron emission
	from a flaring event or thermal emission from the accretion
	disc external edge or from dust surrounding the binary
	system.\\

\section{The plateau state in 1997}
\label{obs97}

   \begin{figure}[!htp]
	\hspace*{-0.4cm}
	\includegraphics[width=9.1cm]{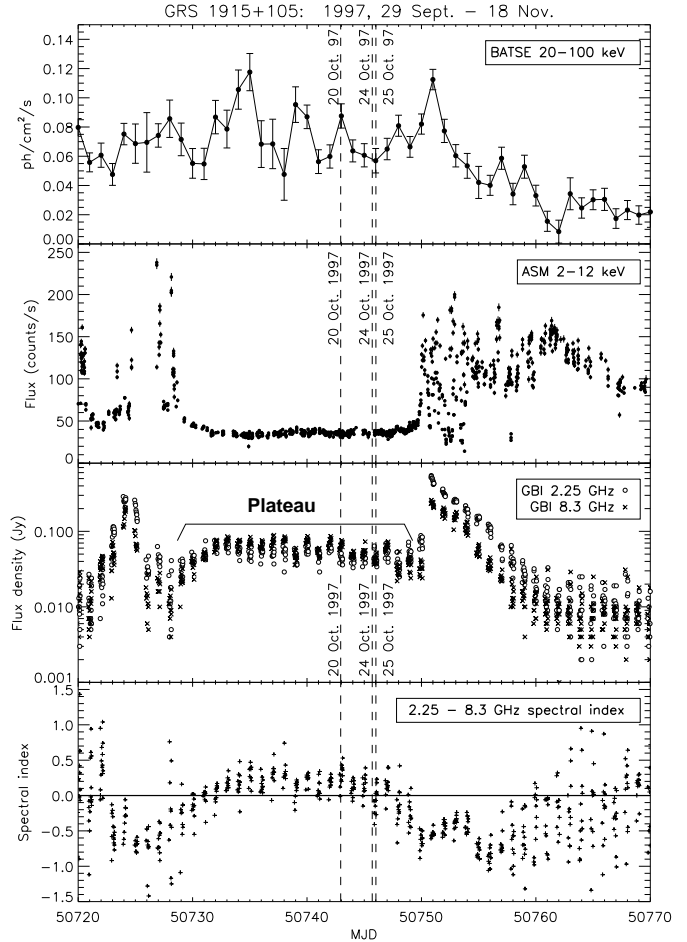}
   	\caption{X-ray and radio flux monitoring of GRS\,1915+105
   	around October 24 1997 (MJD 50745). The upper panels show
   	the photon flux measured by BATSE in the 20--100~keV band and
   	the integrated count rate in the 2--12~keV band by the Rossi
   	X-ray Timing Explorer All-Sky Monitor (RXTE ASM). The lower
   	panels show the corresponding flux densities measured at
   	2.25\,GHz and 8.3\,GHz with the Green Bank Interferometer
   	(GBI) and the spectral index from these two frequencies. On
   	the latter plot the zero value has been marked with a solid
   	line in order to emphasize the optically thick ($\alpha \sim
   	0$) radio emission of the plateau state. The dotted lines mark
   	the begining of our ISOCAM observations on 1997 October. They
   	took place during this noticable plateau state preceding a
   	giant radio outburst with superluminal ejections.}
	\label{figetat97}
    \end{figure}

	\subsection{State of the source}

\begin{figure*}[!htp]
	\includegraphics[width=8.9cm]{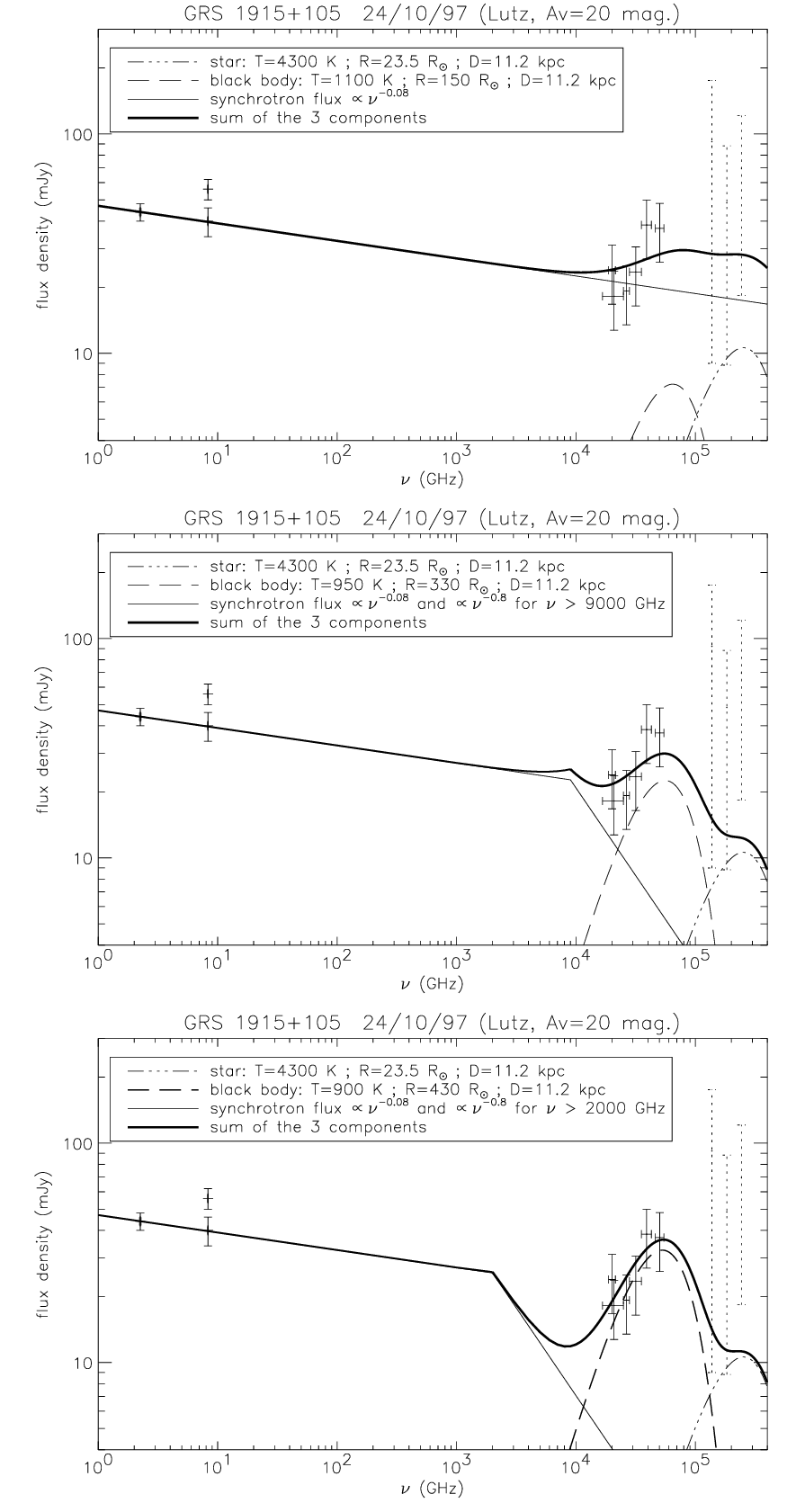}
	\includegraphics[width=8.9cm]{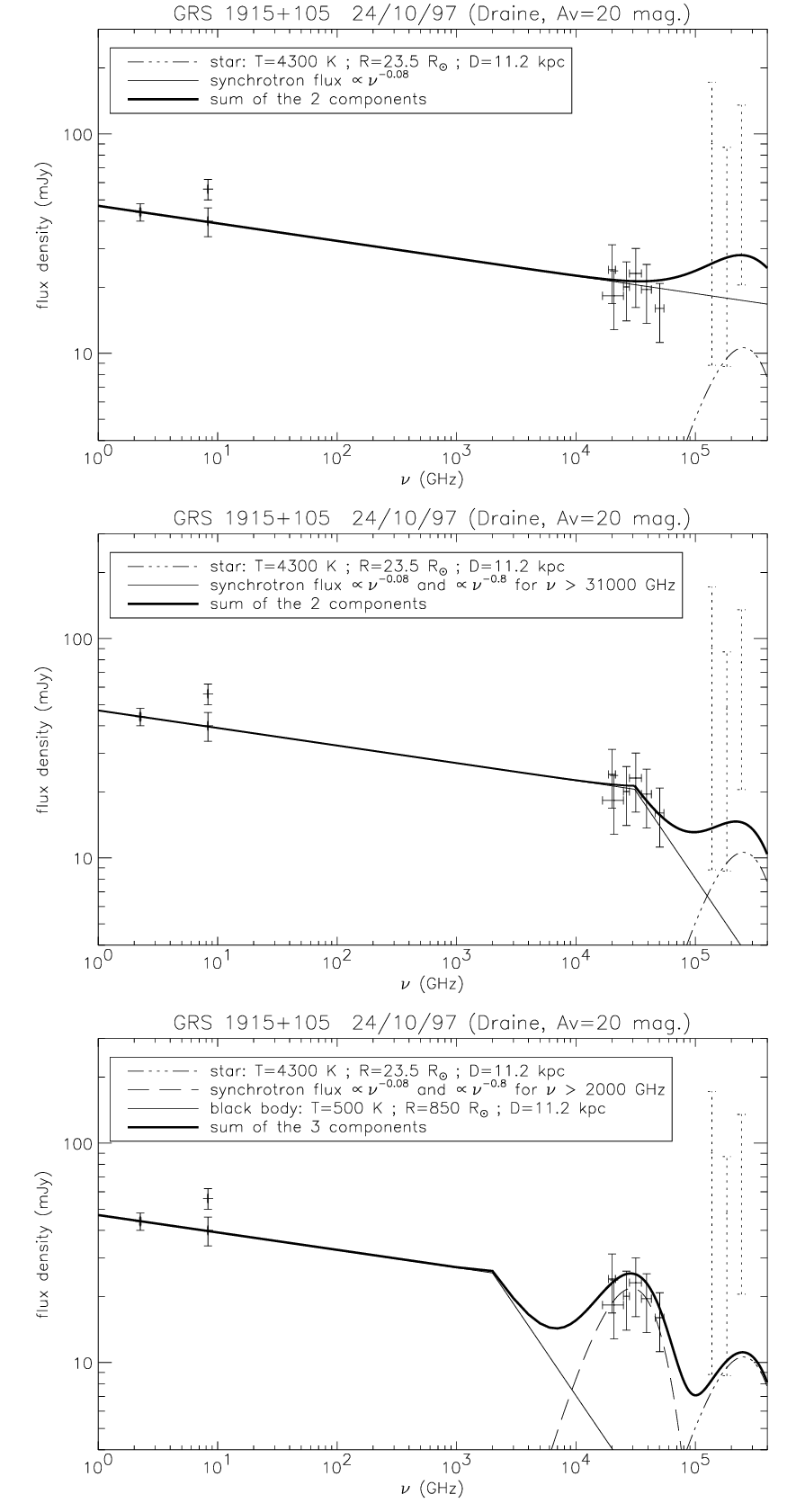}
   	\caption{Radio and ISOCAM spectrum of GRS\,1915+105 in 1997 October 24,
   	dereddened with (left) \citet{lutz96} and (right)
   	\citet{draine89} laws. The dotted lines correspond to the minima
   	and maxima fluxes published in the J (1.25\,$\mu$m $= 2.4 \times
   	10^5$ GHz), H (1.65\,$\mu$m $= 1.81 \times 10^5$ GHz) and K
   	(2.2\,$\mu$m $= 1.36 \times 10^5$ GHz) bands. ISOCAM filters
   	are, from left to right: LW3 ($1.67-2.5 \times 10^4$~GHz), 
	LW9 ($1.875-2.14 \times 10^4$~GHz), 
	LW8 ($2.5-2.8 \times 10^4$~GHz), 
	LW7 ($2.6-3.53 \times 10^4$~GHz), 
	LW6 ($3.53-4.28 \times 10^4$~GHz) 
	and LW4 ($4.61-5.45 \times 10^4$~GHz) (see
   	Table~\ref{tabobsGRS1915}). The GBI 2.25 and 8.3 GHz fluxes
   	observed 14h before (higher level) and 1h (lower level) after
   	the ISOCAM observations are also plotted. These latter GBI values have
   	been used in our tentative fits of the radio to IR fluxes
   	which are plotted as indicated in the legend (see details in
   	the text).}
	\label{figISO97}
    \end{figure*}

	The second set of ISOCAM observations of GRS\,1915+105 took
	place in 1997 October as noted in
	Table~\ref{tabobsGRS1915}. Fig.~\ref{figetat97} shows the
	X-ray and radio flux monitoring of the source during this
	month and the begining of 1997 November. Because of the quasi
	permanent and non predictable variability of the source in the IR,
	we have only plotted in Fig.~\ref{figISO97} flux densities
	measured continuously on October 24 1997 (MJD 50745).

	The GBI flux densities of GRS\,1915+105 
	and their estimated 1\,$\sigma$ errors on this day 
	were at about 2h30 (UT):
	$F_\mathrm{2.25\,GHz} = 44 \pm 4$~mJy, 
	\mbox{$F_\mathrm{8.3\,GHz} = 56 \pm 6$~mJy,} 
	spectral index $\alpha = 0.18 \pm 0.15$ 
	\mbox{($F_\nu \propto \nu^\alpha$)} and at
	about 19h (UT): \mbox{$F_\mathrm{2.25\,GHz} = 44 \pm 4$~mJy,} 
	\mbox{$F_\mathrm{8.3\,GHz} = 40 \pm 6$~mJy,} 
	\mbox{$\alpha = -0.08 \pm 0.18$.} This last
	observation occurred less than one hour after the end of our ISOCAM
	observation so we have used it for our spectral fits (see next
	section) although we have plotted the fluxes corresponding to
	both hours in Fig.~\ref{figISO97}. On that day, X-ray fluxes show a
	low/hard state with a quite steady low RXTE/ASM 2--12~keV level of
	$\simeq 35$~counts/s and an irregular high BATSE 20--200~keV 
	photon flux of $\simeq 0.06$~ph/cm$^2$/s.

	At that epoch, the radio and X-ray monitorings 
	show that GRS\,1915+105 was
	in a very particular state called the {\it plateau} state (see
	Fig.~\ref{figetat97} and Fig.~1 of \citealt{fender99}). This
	state, first described by \citet{foster96} 
	and more
	specified by \citet{fender99}, results in an increase in
	the radio flux density to levels of $\sim$\,50--100~mJy with an
	onset and a decay which can be as short as $\sim 1$~day. The
	{\it plateau} state is characterized by an optically thick radio
	emission ($\alpha \sim 0$) and by a
	hard X-ray emission with a fairly high BATSE 20--100\,keV flux 
	associated to a quasi-stable RXTE/ASM
	2--12\,keV flux. Its duration ranges from one day or two to many
	weeks. The inverted-spectrum radio emission ($\alpha \geq 0$)
	corresponds to powerful self-absorbed quasi-continuous jets
	which were observed by \citet{dhawan00} and have spectral similarities
	with compact jets from other black hole candidate systems in
	similar hard X-ray states \citep{fender01}.
	The {\it plateau} state also corresponds in X-ray to the
	canonical low/hard C state class $\chi$ of \citet{belloni00}
	dominated by a power-law component with little or no disc
	contribution in the X-ray band. More specifically this term
	associated with the presence of compact jets applies to
	$\chi_1$ and $\chi_3$ states as studied by
	\citet{kleinwolt02}, also called radio-loud or radio-plateau
	low/hard states by \citet{muno01} and type II hard steady
	states by \citet{trudolyubov01}.  \citet{vadawale01} have even
	reported detection of synchrotron radiation in the
	0.5--180\,keV X-ray band of RXTE PCA and HEXTE during this
	state, although it has to be confirmed.

	Long (i.e. more than a few days) {\it plateau} states appear to
	precede major optically thin radio flares corresponding to
	superluminal radio ejection events (see also
	\citealt{fenderosc02}). Plateau states are thus of further interest
	in order to understand the release of major superluminal
	outflows.  As shown in Fig.~\ref{figetat97}, our ISOCAM
	observation on October 24 1997 were followed by a
	major flare 5 days later (on MJD 50750) with superluminal
	motion observed by \citet{fender99} with MERLIN. As shown by
	the GBI spectral index, we observed GRS\,1915+105 at the
	transition period from optically thick toward optically thin
	radio emission preparing the major outburst.

	\subsection{Results}

	The ISOCAM flux densities on October 24 1997 show quite different
	spectral energy distributions depending on the dereddening law
	used (\citealt{lutz96} or \citealt{draine89}). As for
	the 1996 observations, we have fitted the mid-IR emission with
	very simple models given the large error bars on the ISOCAM
	flux densities ($\sim$\,30\%).  In Fig.~\ref{figISO97} we have plotted the
	contribution of the donor star with the same parameters:
	$\mathrm{T}=4300$~K, $\mathrm{R}=23.5$~R$_\odot$
	(dashed-dotted line).

	As the source was in a plateau low/hard state, we have fitted
	the radio emission with a power law representing the
	synchrotron radiation from the compact jets (solid thin
	line). But whether in this state the synchrotron component
	extends to the IR range, as has been 
	asserted by \citet{fender01}, is questioned. This is why we
	have tried the three following cases: (1) no break in the
	synchrotron emission, (2) a break occuring close to the mid-IR
	range (respectively at 9000\,GHz and 31\,000\,GHz for the
	\citealt{lutz96} and \citealt{draine89} laws) so that the
	synchrotron emission contributes to the mid-IR flux densities,
	and (3) a break occuring before the mid-IR range so that the
	synchrotron emission is negligible at those frequencies. The
	optically thick part of this synchrotron component has been
	modeled using the GBI flux densities measured only 1\,h after
	ISOCAM thus with the corresponding spectral index
	$\alpha=-0.08$. The slope of this synchrotron emission after
	the break ($\alpha =-0.8$) has been fitted according to the
	ISOCAM flux densities in the LW4, LW6 and LW7 filters
	dereddened with the \citet{draine89} law.

	Fig.~\ref{figISO97} shows that case (1) can fit flux densities
	dereddened with the \citet{draine89} law as they are grouped
	around the same level ($\sim 20$~mJy) and the $\alpha=-0.08$ 
	spectral index perfectly matches the mid-IR flux level despite 
	the large error on this index ($\pm 0.18$). 
	But case (1) cannot fit flux
	densities dereddened with the \citet{lutz96} law where an
	additional component is clearly needed for the LW6 
	($3.53-4.28 \times 10^4$~GHz = 7--8.5~$\mu$m) and LW4 
	($4.61-5.45 \times 10^4$~GHz = 5.5--6.5~$\mu$m) filters. Case
	(2) is satisfactory for flux densities dereddened with the
	\citet{draine89} law if the break occurs in the mid-IR range
	(31\,000\,GHz = 9.7\,$\mu$m). In all the other cases an
	additional component is needed. We then have used a black body
	model (long-dashed curves) standing for thermal emission from
	heated dust, with the following parameters: 
	for the flux densities
	dereddened with the \citet{lutz96} law $\mathrm{T}=1100$~K and
	$\mathrm{R}=150$~R$_\odot$ (1), $\mathrm{T}=950$~K and
	$\mathrm{R}=330$~R$_\odot$ (2), $\mathrm{T}=900$~K and
	$\mathrm{R}=430$~R$_\odot$ (3); for the flux densities dereddened
	with the \citet{draine89} law $\mathrm{T}=500$~K and
	$\mathrm{R}=850$~R$_\odot$ in case (3).  Note that these
	values must only be considered as orders of magnitude since
	the black body model is a very simple approximation for dust
	emission.

	We can summarize the results of our fits, represented in
	Fig.~\ref{figISO97} by the sums of the components in thick
	solid lines, as the following. The synchrotron emission
	extended to the mid-IR range can account for the
	flux densities in this wavelength range in the case of
	dereddening with the \citet{draine89} law. In this latter case
	the mid-IR flux densities can also be explained by thermal
	emission from possible heated dust. Dereddening with the
	\citet{lutz96} law implies the need for a third component
	accounting for most of the mid-IR emission, possibly thermal
	emission from hot dust.

	\subsection{Interpretation / discussion}

	\subsubsection{Dust emission ?}

	If the mid-IR emission is due to hot dust, then one has to
	explain its presence and its heating mechanism.

	First we question if the source of this dust could be the K-M
	giant donor star. Such a star has left the main sequence
	phase to evolve toward the AGB one. When it is isolated,
	its atmosphere expands and their mass loss increases to 
	\mbox{$\sim 10^{-9} - 10^{-8}$~M$_\odot$.yr$^{-1}$} during
	this giant phase. 
	As GRS\,1915+105 is a binary system,
	the atmosphere expansion is limited by the
	Roche lobe size, and in this case the star has very likely
	reached its maximum size and has entered in the Roche lobe
	overflow phase with an accretion process characteristic of
	low-mass X-ray binaries. In order to produce dust, the
	material lost by the companion should escape from the black
	hole gravitational attraction, so the mass loss should
	exceed the mass accretion rate. This is unlikely as the latter
	has been evaluated by \citet{belloniIR00} to $\sim 10^{-8} -
	10^{-7}$~M$_\odot$.yr$^{-1}$ during IR quasi-periodic
	oscillations even if \citet{belloni00} has estimated it to be very
	low during the plateau intervals.
	But even if the donor mass loss was high enough, the wind
	from a giant star is supposed to be composed of atomic
	elements and not of molecular ones which are on the other hand
	needed for dust production. Moreover the latter is usually not
	enabled with isolated giant stars since they are too hot. Thus
	it is unlikely that the donor star is a source of dust.

	The dust could come from an external source of the binary
	system: the bright star seen in our ISOCAM images (see
	section~2.1.2). This bright star is probably in the AGB phase
	according to the study of \citet{felli00}. In order to stay consistent with this study, we took the magnitudes of this star from the ISOGAL survey
	(http://www-isogal.iap.fr/): 
	$\simeq 7.99$~mag at 7\,$\mu$m (LW2) and 
	$\simeq 7.92$~mag at 15\,$\mu$m (LW3)
	giving [7]~$\simeq 7.43$~mag and
	[15]~$\simeq 7.62$~mag when dereddened with $A_V =20$~mag and 
	the relations $A_7 = 0.028 A_V$ and $A_{15} = 0.015 A_V$ as in
	\citet{felli00}, so [7]--[15]~$\simeq
	0.19$~mag which corresponds to an AGB class in the figure~1 of
	\citet{felli00}. If we assume that this star is located at the
	same distance from the Earth as GRS\,1915+105, then the 18$''$
	angular separation between this AGB star and GRS\,1915+105
	corresponds to $\sim 0.58$~pc at 6.6~kpc and $\sim 0.98$~pc at
	11.2~kpc. An AGB star undergoes much mass loss (till
	$\sim 10^{-4}$~M$_\odot$.yr$^{-1}$) and can eject material in
	its surroundings as far as 1\,pc. But at such a distance only
	little material is provided and so it may not be able to
	account for the mid-IR emission from GRS\,1915+105.

	On the other hand, if we assume that this AGB star is much
	closer than GRS\,1915+105, its surrounding nebula caused by
	the mass loss probably contaminates the line of sight to this
	source. This might explain the anomalous Si and Fe abundances
	deduced by \citet{lee02} from Chandra spectra of
	GRS\,1915+105, instead of dust that they suggested among other
	possibilities.\\
 
	The problem is also to explain the high temperatures (900 to
	1100~K) of the dust and the distances found with
	our fits.  The distance, 150 to 850~R$_\odot$, represents 6 to
	35 times the chosen radius of the giant star
	(23.5~R$_\odot$) which is constrained by the gravitational
	interactions with the black hole (R $\leq 21 \pm 4$~R$_\odot$
	the Roche lobe size estimated by
	\citealt{greiner01Nat}). These parameters would not be
	surprising for dust surrounding some AGB stars, as shown by
	\citet{danchi94} who found M stars with inner dust shells very
	close to the photospheres of the stars ($3-5$ stellar radii)
	and at high temperature ($\sim$\,1200~K). But for a giant star
	with a magnitude of K=14.5--15.0 as inferred by
	\citet{greiner01} for GRS\,1915+105, the
	luminosity of the this donor star  illustrated by the
	black body in dashed-dotted line in Fig.~\ref{figISO97}
	is not high enough to induce the luminosity of the black body
	representing the possible dust on our fits.  Then the dust
	heating in GRS\,1915+105 is unlikely to be carried out by the
	donor star.

	Another possible origin of the dust heating might be the high
	energy activity of GRS\,1915+105. As a consequence, changes in
	the observed temperature of this dust between our fits in 1996
	and in 1997 (when mid-IR flux densities is higher) might be due to
	changes in the X-ray activity.  Thus the low/hard radiations
	coming from the jets or from Compton reprocessing of the soft
	disc photons by the corona would be more efficient than
	high/soft emission from the accretion disc.
	However, the efficiency of dust heating by X-rays is
	questioned by \citet{vanpar94} who discussed the origin of the
	10\,$\mu$m emission from the low-mass X-ray binary 
	\object{GRO\,J0422+32}. They found, from the study of \citet{voit91} on
	X-ray irradiation of interstellar grains, that this mid-IR
	emission cannot be explained by the X-ray heating of dust,
	which leads to flux densities at least two orders of magnitude below
	the observed value, unless the system is surrounded by a highly
	non-standard interstellar medium. \\

	Note that the hypothesis of IR emission due to dust
	surrounding GRS\,1915+105 was initially suggested by
	\citet{mirabel96} from the observation of an IR flare occuring
	2 days after a radio flare and corresponding to a reddening
	between the J and K band observations taken
	quasi-simultaneously. However, at that epoch the
	$\sim$~30~minutes cycles of flares were not known and the
	observations in each near-IR band took about 10~minutes. As a
	consequence these flux densities are time average ones and
	are not reliable for a spectral
	indication. It is very likely that on August 15 1995, the
	observed flare and reddening of GRS\,1915+105 by
	\citet{mirabel96} were due to a $\sim$~30~minutes flaring
	event since it happened 5 days after a giant radio outburst,
	now considered as the sign of superluminous ejections, and
	since \citet{fender99,fenderosc02} have observed these
	quasi-periodic oscillations in radio during the decay of such
	giant outbursts.

	Thus if we draw the parallel with symbiotic stars (for a
	review see \citealt{mikolajewska02}) GRS\,1915+105 belongs to
	the $\sim$\,80\% of S-type (stellar) systems containing a
	normal giant (the remaining $\sim$\,20\% are D-type containing
	Mira-type variable with a dust shell). However it is a very
	peculiar one with a black hole instead of a white dwarf, with
	a short binary period and orbital separation leading to Roche
	lobe overflow instead of the high predominance of
	wind-accretion in usual symbiotic stars.

	\subsubsection{Other: synchrotron or free-free emission}

	Thermal emission from dust seems so complicated to explain
	that another solution is more likely. As we observed
	GRS\,1915+105 during the {\it plateau} state, which is known to
	correspond to the presence of a compact jet observed by
	\citet{dhawan00}, the high-frequency extension of the radio
	synchrotron emission to the mid-IR logically ensues.  This
	hypothesis was first suggested by \citet{fender01}
	gathering the radio data of \citet{pooleyfender97}
	with the only K-band flux from \citet{bandyo98} during the {\it
	plateau} state of 1996 August. However these observations were
	not simultaneous and Fig.~\ref{figISO97} suggests that the
	K-band flux probably corresponds not only to synchrotron
	emission (if at all) but that the giant companion contribution
	may not be negligable in this band. In our study, using
	the \citet{draine89} law, the extension of the radio
	synchrotron emission can account for the totality of the
	mid-IR emission measured in several bands.  It may
	be the first time that we have evidence of synchrotron
	emission in mid-IR for GRS\,1915+105 in the {\it plateau}
	state.
	In this case, GRS\,1915+105 is similar to other black hole
	candidate systems in hard X-ray states showing powerful
	self-absorbed quasi-continuous jets \citep{fender01} with
	synchrotron emission extending to the near-IR or optical ranges
	such as \object{GX\,339$-$4} \citep{corbelfender02},
	\object{XTE\,J1118+480} \citep{markoff01,fender111801} and
	\object{XTE\,J1550$-$564} \citep{corbel01}.

	The synchrotron emission in the mid-IR range seems very likely
	but we cannot rule out another possibility corresponding to
	the flat spectrum obtained with the \citet{draine89} law:
	optically thin free-free emission from an X-ray driven wind
	from the accretion disc. This kind of emission was suggested
	by \citet{vanpar94} for GRO\,J0422+32 to explain their
	observed 10~$\mu$m flux, although \citet{fender01}
	suggests that this detection was also synchrotron emission
	from the compact self-absorbed jet. The presence of a strong
	radiatively driven wind was also suggested by \citet{lee02} to
	explain ionized features in the X-ray spectrum of
	GRS\,1915+105, although evidence of this might be difficult to
	observe. Such a wind may be formed by X-ray heating of the
	disc for X-ray luminosities in excess of a few percent of the
	Eddington limit \citep{begelman83} which is probably the case
	for GRS\,1915+105 \citep{king02}.  Infrared free-free emission
	from a disc wind may be common for X-ray binaries as it has
	already been observed in \object{Cygnus\,X-3}
	\citep{fendercygx399,koch02} and possibly in \object{SS\,433}
	(\citealt{wynn79}, \citealp[ in\ prep.]{fuchs}).

	Note that either synchrotron or free-free emission corresponds
	to the flat spectrum resulting from dereddening with the
	\citet{draine89} law. The latter appears more likely than 
	the \citet{lutz96} law since it leads to satisfactory
	solutions to account for the mid-IR emission without the need
	for an additional component as thermal emission from dust
	(which is not easy to explain). Thus our 1997 observation
	enables us to prefer and to recommend the use of the
	\citet{draine89} law to deredden the mid-IR observations of
	GRS\,1915+105.\\

\section{Conclusions}

\label{concl}
	We have presented for the first time mid-IR observations 
	of GRS\,1915+105 obtained with ISOCAM.
	The near-IR and ISOCAM observations during the high/soft state 
	in 1996 are difficult 
	to interpret since they possibly took place during a flaring 
	event with a too fast time variability compared to the response
	of the mid-IR detectors. 
	The emission from the giant donor star alone
	cannot account for the near and mid-IR emission and one or
	several additional components are needed.

	The other ISOCAM observation took place in October 1997, while
	GRS\,1915+105 was in the peculiar {\it plateau} state
	characterized by a flat or slightly inverted radio spectrum due
	to the synchrotron emission from compact jets.  This state is of
	further interest as it preceded a giant radio flare
	corresponding to a major superluminal ejection event.  We
	showed for the first time observations during a plateau state
	with quasi-simultaneous radio and mid-IR measurements.  We
	discussed the different possibilities concerning the
	latter.  Due to the large error-bars neither solution can be
	totally excluded but the presence of dust and its inferred
	temperature is unlikely. 
	Synchrotron
	emission can account for the flat broadband spectrum from the
	radio to the mid-IR range.
	Thus the mid-IR emission during the {\it plateau} state of
	GRS\,1915+105 is very likely to be the extension of the radio
	synchrotron emission, although optically thin free-free
	emission of an X-ray driven-wind from the accretion disc
	cannot be excluded.  The 1997 observation enables us to choose
	between the two dereddening laws: the one from
	\citet{draine89} giving a flatter spectrum in the mid-IR
	appears more plausible than the \citet{lutz96} law.
	


\begin{acknowledgements}
	We warmly thank Anthony Jones, Vivek Dhawan, Claude Chapuis,
	Jerome Rodriguez and Tibault Le Bertre for very informative
	and useful discussions. We also thank Guy Pooley for kindly
	providing RT monitoring flux densities.  The ISOCAM data presented in
	this paper were analysed using ``CIA", a joint development by
	the ESA Astrophysics Division and the ISOCAM Consortium. The
	ISOCAM Consortium was led by the ISOCAM PI, C. Cesarsky,
	Direction des Sciences de la Matière, C.E.A, France.  We thank
	the organisations providing the public data that we have used:
	quick-look results provided by the ASM/RXTE team, high-level
	data products from the CGRO BATSE instrument which were
	generated by the BATSE Instrument Team at the Marshall Space
	Flight Center (MSFC) using the Earth occultation technique,
	and GBI monitoring programme. The Green Bank Interferometer is
	a facility of the National Science Foundation operated by the
	National Radio Astronomy Observatory, in support of USNO and
	NRL geodetic and astronomy programs, and of NASA High Energy
	Astrophysics programs.  Y.F. is supported by a CNES external
	post-doctoral fellowship.
\end{acknowledgements}

\bibliographystyle{aa}
\bibliography{refgrs}




\end{document}